\documentclass[notitlepage,twocolumn,superscriptaddress,amsmath,amssymb,aps,pra]{revtex4-2}
\usepackage{amsmath,amssymb,graphicx,mathtools,dsfont,ulem,bbold}
\usepackage[dvipsnames]{xcolor}
\usepackage[T1]{fontenc}
\usepackage{hyperref}
\usepackage{soul}
\usepackage{bibunits}
\hypersetup{citecolor=red,colorlinks=true,urlcolor=blue}

\newcommand{\ket}[1]    {|#1 \rangle}

\newcommand{\trs}[1]    {{\rm Tr}\big[ #1 \big]}
\newcommand{\av}[1]    {\langle #1 \rangle}

\newcommand{\modsq}[1]    {\left| #1 \right|^2}
\newcommand{\modsqs}[1]    {\big| #1 \big|^2}

\newcommand{\En}{\mathcal E_N}
\newcommand{\Q}{\mathcal Q}

\usepackage{diagbox} 

\usepackage{hhline}

\begin{document}

\preprint{  }

\title{Entanglement classification and \emph{non-k}-separability certification via Greenberger-Horne-Zeilinger-class fidelity
}% Force line breaks with \\
% \thanks{A footnote to the article title}%
%AUTHORS
 
\author{Marcin Płodzień}
\affiliation{ICFO-Institut de Ciencies Fotoniques, The Barcelona Institute of Science and Technology, 08860 Castelldefels (Barcelona), Spain}

\author{Jan Chwedeńczuk}
\affiliation{Faculty of Physics, University of Warsaw, ul. Pasteura 5, 02-093 Warszawa, Poland}

\author{Maciej Lewenstein}
\affiliation{ICFO-Institut de Ciencies Fotoniques, The Barcelona Institute of Science and Technology, 08860 Castelldefels (Barcelona), Spain}
\affiliation{ICREA, Passeig Lluis Companys 23, 08010 Barcelona, Spain}
 
\author{Grzegorz Rajchel-Mieldzioć}
\affiliation{NASK National Research Institute, ul. Kolska 12, 01-045 Warszawa, Poland}
\affiliation{ICFO-Institut de Ciencies Fotoniques, The Barcelona Institute of Science and Technology, 08860 Castelldefels (Barcelona), Spain}

\begin{abstract}
 
Many-body quantum systems can be characterised using the notions of  \emph{k}-separability and entanglement depth. A quantum state is \emph{k}-separable if it can be expressed as a mixture of \emph{k} entangled subsystems, and its entanglement depth is given by the size of the largest entangled subsystem. In this paper we propose a multipartite entanglement measure that satisfies the following criteria: (i) it can be used with both pure and mixed states; (ii) it is encoded in a single element of the density matrix, so it does not require  knowledge of the full spectrum of the density matrix; (iii) it can be applied to large systems; and (iv) it can be experimentally verified. The proposed method allows the certification of \emph{non-k}-separability of a given quantum state. We show that the proposed method successfully classifies three-qubit systems into known stochastic local operations and classical communication (SLOCC) classes, namely bipartite, \mbox{W-,} and GHZ-type entanglement. Furthermore, we characterise the \emph{non-k}-separability in known nine SLOCC classes of four-qubit states,  absolutely maximally entangled states for five and six qubits and for arbitrary size qubit Dicke states. 

\end{abstract}

% To cite
% https://arxiv.org/abs/2406.07274
 
\maketitle

%\tableofcontents

\section{Introduction}
 
One of the important consequences of the quantum description of physical systems is the notion of  entanglement~\cite{horodecki2009entanglement} which is the necessary resource for the Einstein-Podolsky-Rosen steering \cite{uola2020steering} and the Bell correlations \cite{bell,brunner2014bell,bell_rmp}. None of these correlations has a classical counterpart. 
The highlighted role of entanglement requires a systematic  characterisation and classification of quantum states in terms of some entanglement measure \cite{PhysRevLett.78.2275,Horodecki2001,Plenio_2007}.
Nevertheless, the tools for entanglement classification are usually difficult to implement in realistic experimental settings and are not universal due to the large number of entanglement classes~\cite{Guhne_2009}. 
The problem of assigning  a multipartite state to some entanglement class is hard even for three qubits~\cite{Bengtsson_2006,Sabin_2008,e20100745,DeChiara_2018,PhysRevA.92.042329,Srivastava_2024,Xie2024-du}, and is even more complicated for four qubits, where  the number of entanglement classes is already infinite~\cite{Verstraete_2002,10.1063/1.3511477,10.1063/1.4946895}.
Hence the variety of multipartite entanglement measures, such as  based on local~\cite{PhysRevA.95.022301} or on the collective measurements~\cite{PhysRevA.69.052327}.
Research has focused on limiting the experimental effort~\cite{PhysRevLett.117.210504} and
recently a method called \emph{entanglement detection length} has been proposed to witness many-body entanglement with a minimal number of observables~\cite{shi2024entanglement}.
A technique has been developed for detecting multipartite entanglement based on a constant number of measurements, independent of the number of qubits, suitable for the Greenberger-Horne-Zeilinger (GHZ)  states, cluster- or Dicke states~\cite{PhysRevLett.112.155304, PhysRevLett.103.020504}. 
The urgency for a versatile entanglement certification method is underlined by the fact that the Dicke states
have been created in experiments with few qubits in different configurations such as photonic systems~\cite{PhysRevLett.103.020504, PhysRevA.83.013821, PhysRevLett.103.020503, Zhao:15, Wang2016, Mu:20}, 
ultracold atoms~\cite{PhysRevLett.112.155304,doi:10.1126/science.1208798,Hamley2012,doi:10.1073/pnas.1715105115} and quantum circuits~\cite{ 9951196, Narisada2023}.
A Hilbert-Schmidt distance between a given state and the closest separable state is another important entanglement witness~\cite{PhysRevA.102.012409}. 
The other methods to distinguish entanglement of quantum states use local unitary optimisations ~\cite{Rudnicki_2011,Rudnicki_2012}. The next measures are entanglement of formation \cite{10.5555/2011326.2011329,PhysRevLett.80.2245,Luo2008,Patrick_M_Hayden_2001,Adam_W_Majewski_2002,FEI2003333,Zhao_Hui_2007,Gao2008,PhysRevA.102.042408}, quantum discord \cite{https://doi.org/10.1002/andp.200051211-1204,L_Henderson_2001,PhysRevLett.88.017901,PhysRevA.77.042303,PhysRevA.79.042325,PhysRevLett.105.030501,PhysRevA.88.014302,PhysRevLett.112.210401,Bera_2018}, and concurrence \cite{PhysRevLett.78.5022,10.1063/1.2795840,PhysRevA.105.052441,Bhaskara2017,PhysRevA.64.042315,PhysRevA.86.062303,Mintert_2005,Ohnemus_2023}.

A separable class of witnesses to multipartite entanglement is related to quantum metrology \cite{Toth_2014}. The prominent example is the quantum Fisher information (QFI), 
which sets the upper limit for the precision of parameter estimation~\cite{braunstein1994statistical}.
Broadly speaking, an increasing value of the QFI implies increasing depth of multipartite entanglement, decreasing $k$-separability, and a better precision of a quantum sensor taking this state as  input~\cite{hyllus2012fisher,toth2012multipartite,PhysRevLett.126.080502}. 
The QFI can be difficult to calculate
and even harder to measure as it requires full information about the spectrum of a many-body state. However other related quantities are easier to access, 
such as the classical Fisher information, the spin-squeezing,
or other witnesses based on two-body correlators that carry some information about the multipartite entanglement~\cite{kitagawa1993squeezed,wineland1994squeezed,PhysRevLett.95.120502,PhysRevA.74.052319,esteve2008squeezing,PhysRevA.84.022107,smerzi_ob,DeChiara2011,
PhysRevA.67.022112, PhysRevA.97.020301, PhysRevLett.107.240502, PhysRevLett.86.4431, Tura2014,Tura2014a,Tura2015,PhysRevA.100.032307,Baccarri2019,PhysRevA.79.042334,PhysRevLett.99.250405, PhysRevA.95.032330,PhysRevA.89.032307, Tura2014, marty2017multiparticle, Vitagliano_2017, PhysRevLett.127.010401, PRXQuantum.2.030329, PhysRevA.100.032307, Frowis2017,  PhysRevLett.123.100507, ROIK2022128270, PhysRevLett.116.093602, MullerRigat2023certifyingquantum, Fadel_2023, Vitagliano2023numberphase, PRXQuantum.2.030329,PhysRevLett.131.070201,Poggi2024measurementinduced}.

Another method is based on the measurement of macroscopic 
observables such as magnetic susceptibility \cite{Wiesniak_2005}, heat capacity \cite{PhysRevB.78.064108}, energy \cite{brukner2004macroscopic, PhysRevA.70.062113, PhysRevA.71.010301}, mean  spin and spin correlations \cite{PhysRevA.72.032309, Frydryszak2017, KUZMAK2020126579}, with magnetic resonance spectroscopy \cite{PhysRevA.100.022330,Lazarev2020}, entropic correlations \cite{PhysRevResearch.2.043152}, out-of-time-order correlations \cite{Lewis-Swan2019,PhysRevA.106.042429,PRXQuantum.2.020339} or the energy of 
a thermal state in equilibrium~\cite{PhysRevResearch.5.013158}.
Next, there is a family of methods based on SLOCC (stochastic local operations and classical communications) and tensor properties \cite{PhysRevA.62.062314, PhysRevA.109.032424, PhysRevLett.110.030501,10.5555/2011734.2011739,PhysRevLett.111.110502,PhysRevA.84.062306,Zangi_2017,PhysRevLett.111.060502,Palazuelos2022genuinemultipartite}. These methods can generally be divided into subgroups~\cite{Guhne_2009}: eigenvalue criteria ~\cite{Horodecki_1996,Peres_1996,Chen_2003,Rudolph_2005,PhysRevLett.106.190502}, entanglement witnesses~\cite{Guhne_2006, TERHAL2002313, PhysRevA.63.044304, Acin_2001_mixed, PhysRevLett.92.087902, Horodecki_2003,PhysRevA.77.062304,PhysRevX.8.021072,PhysRevLett.127.040401,Rico_2024}, geometric and robustness criteria~\cite{Vidal_1999,PhysRevA.68.042307,10.1063/1.1497700,doi:10.1142/S1230161222500111,9605268,Guo_2022,quantum2010004,PhysRevLett.127.040403,PhysRevResearch.4.023059}, and operational measures~\cite{Klyachko2006,Plenio_2007,PhysRevLett.115.150502}.  Entanglement has been also classified with the help of mathematical tools from group theory \cite{PhysRevA.74.022318}, topology \cite{PhysRevD.107.126005}, and algebraic geometry \cite{PhysRevResearch.2.043003}.
Finally, in recent years the problem of entanglement classification in many-body systems has been approached from a machine learning point of view \cite{PhysRevA.98.012315, Harney_2020, Harney_2021,Chen_2021, s22186767, ayachi2022general, https://doi.org/10.1002/qute.202200025, Vintskevich_2023, Giordano_2022, Asif2023, Sanavio2023, pawlowski2023identification, doi:10.1126/sciadv.add7131, palmieri2023enhancing, PhysRevA.109.022405, dawid2023modern, Wiesniak2023, urena2024entanglement,brunner2024,fuchs2024}.

Here we propose a method for detecting and classifying many-body entanglement using a correlator  originally proposed for detecting Bell correlations
and the Einstein-Podolsky-Rosen steering~\cite{cavalcanti2007bell,he2010bell,cavalcanti2011unified,spiny.milosz,PhysRevLett.126.210506}. 
Among the advantages of this method is the possibility to quantify the entanglement with a single element of the density matrix, 
which is directly accessible in an experiment without the need for full tomography~\cite{garttner2017measuring,PhysRevLett.120.040402,PRXQuantum.2.010307}. 
It has already been shown to detect the entanglement depth in the multipartite setting~\cite{spiny.milosz,PhysRevA.102.013328, PhysRevLett.129.250402,PhysRevResearch.6.023050}.
This correlator belongs to a broader family considered in Ref.~\cite{zukowski2002bell}, where a general framework for certifying many-body Bell correlations was formulated. 
Its other variants include the Mermin form of Bell inequalities~\cite{Mermin_1990, PhysRevLett.100.200407,10.1119/1.12594},
the $N$-body inequalities~\cite{zukowski2002bell, cavalcanti2007bell, Reid2012, cavalcanti2011unified} or the Ardehali inequalities~\cite{Ardehali_1992}. 
Some of these inequalities have been used to distinguish W and GHZ states~\cite{Brunner_2012}.
Recently, several other techniques have been developed that tackle the problem of certifying entanglement without using all possible $N$-body correlations~\cite{Friis_2018,Frerot_2022}.

Our method allows SLOCC classification of three- and four-qubit entanglement classes for both pure and mixed states. 
We use the proposed setup to characterise the $k$-separability of all remaining qubit absolutely maximally entangled (AME) states, and to classify the $k$-separability of Dicke states of any number of qubits. All characterisations are performed by inspecting the single element of the density matrix,  corresponding to the GHZ coherence in the optimally chosen basis.

The paper is structured as follows: 
In Sec.~\ref{sec:Methods} we introduce the framework used for characterisation and classification of  \emph{non-k}-separability. Next, in Sec.~\ref{sec:Results} we use the proposed method to characterise \emph{non-k}-separability in many-body qubit systems. In Sec.~\ref{sec:Results_3} we focus on three-qubit quantum states and show its relation with SLOCC classification. In  Sec.~\ref{sec:Results_4} we characterize \emph{non-k}-separability of four-qubit states from nine SLOCC classes; in Sec.~\ref{sec:Results_AME} we characterise Absolutely Maximally Entangled (AME) states and 
in Sec.~\ref{sec:Results_Dicke} we characterise $k$-separability of Dicke states for any number of qubits and any magnetisation sector. In Sec.~\ref{sec:Measurement} we comment on the protocol for measuring \emph{non-k}-separability. 
Finally, we conclude in Sec.~\ref{Sec:Conclusions}. Some additional details are presented in Appendices. 

\section{Methods}\label{sec:Methods}

\subsection{Many-body correlator}
Consider a many-body system containing $N$ parties.  Each of $N$ parties can measure a pair of observables $\sigma^{(k)}_x$ and $\sigma^{(k)}_y$,  each yielding binary outcomes, i.e., $\sigma_{x/y}=\pm1$,
with $k\in\{1\ldots N\}$. 
At this point, it is not assumed that these observables are quantum.
The correlation function (also referred to as the correlator) of these results is here defined as
\begin{align}\label{eq.el}
  \En=\modsq{\av{\sigma_+^{(1)}\dots\sigma_+^{(N)}}},
\end{align}
with $\sigma_+^{(k)}=1/2(\sigma_x^{(k)}+i\sigma_y^{(k)})$. This correlation is consistent with the local and realistic theory if the average can be expressed in terms of an integral
over the random (\emph{hidden}) variable $\lambda$ distributed with a probability density $p(\lambda)$ as follows
\begin{align}\label{eq.lhv}
  \av{\sigma_+^{(1)}\dots\sigma_+^{(N)}}=\int\!\!d\lambda\,p(\lambda)\prod_{k=1}^N\sigma_+^{(k)}(\lambda).
\end{align}
Using the Cauchy-Schwarz inequality for complex integrals together with the fact that $\modsq{\sigma_+^{(k)}(\lambda)}=1/2$ for all $k$,  we see that
the ${\cal E}_N$ is bounded from above:
\begin{align}
  \En&=\modsq{\int\!\!d\lambda\,p(\lambda)\prod_{k=1}^N\sigma_+^{(k)}(\lambda)}\nonumber\\
  &\leqslant\int\!\!d\lambda\,p(\lambda)\prod_{k=1}^N\modsq{\sigma_+^{(k)}(\lambda)}=2^{-N}.
\end{align}
We thus conclude that
\begin{equation}
{\cal E}_N\leqslant2^{-N}
\end{equation}
is the $N$-body Bell inequality, because its violation defies the postulates of local realism expressed in Eq.~\eqref{eq.lhv}. 

The provided inequality is valid for systems that yield binary outcomes of local, 
single-particle observables - such conditions are realized in the case of the $N$-qubit  quantum system described by density matrix $\hat{\varrho}$. The quantum-mechanical equivalent of Eq.~\eqref{eq.el} is defined as
\begin{align}\label{eq.elq}
  {\cal E}=\modsq{{\rm Tr}\bigg[{\hat{\varrho}\bigotimes_{k=1}^N\hat\sigma_+^{(k)}}\bigg]}.
\end{align}
Assuming quantum-mechanical behavior of each parties of the system, i.e., $\modsqs{\langle \hat{\sigma}_+^{(k)}\rangle}\le 4^{-1}$, we obtain that for a separable state  of $N$ qubits the following holds
\begin{align}\label{eq.ent.l}
  {\cal E}\leqslant 4^{-N}.
\end{align}
Based on the above considerations, to witness and classify the multi-qubit entanglement of an $N$-qubit density matrix  $\hat\varrho$, we use the generalized $N$-body correlator 
\begin{align}\label{eq.eqn}
  \Q= \log_4\big[4^N\modsqs{\trs{\hat\varrho\,\bigotimes_{k=1}^N\hat\sigma_{+,n_k}^{(k)}}}\big],
\end{align}
where $\hat\sigma_{+,n_k}^{(k)}$ is the single-qubit operator that is rising the projection of the spin along the axis $n_k$, i.e., 
\begin{align}
  \hat\sigma_{+,n_k}^{(k)}=\frac12(\hat{\sigma}^{(k)}_{m_k}+ i\hat{\sigma}^{(k)}_{l_k}).
\end{align}
where $\vec r_k=(n_k,m_k,l_k)$ are three  mutually orthogonal components of a Bloch vector. 
Axis $n_k$ defines the rotated version of the operator $\hat{\sigma}_+^{(k)}=1/2(\hat{\sigma}_x^{(k)} +i\hat{\sigma}_y^{(k)})$ after the action of local unitary matrices, i.e., $ \hat\sigma_{+,n_k}^{(k)} = \hat{R}_k^\dagger(\vec{\theta}_k) \hat{\sigma}_+^{(k)} \hat{R}_k(\vec{\theta}_k),$ where $\hat{R}_k(\vec{\theta}_k) = e^{-i\vec{\theta}_k\cdot \vec{\sigma}_k}$,   $\vec{\theta}_k = \{\theta^{(k)}_x, \theta^{(k)}_y, \theta^{(k)}_z\}$, $\vec{\sigma}_k = \{\hat{\sigma}_x^{(k)}, \hat{\sigma}_y^{(k)}, \hat{\sigma}_z^{(k)} \}$.

\begin{figure}[t!]
    \centering
    \includegraphics[scale=0.34]{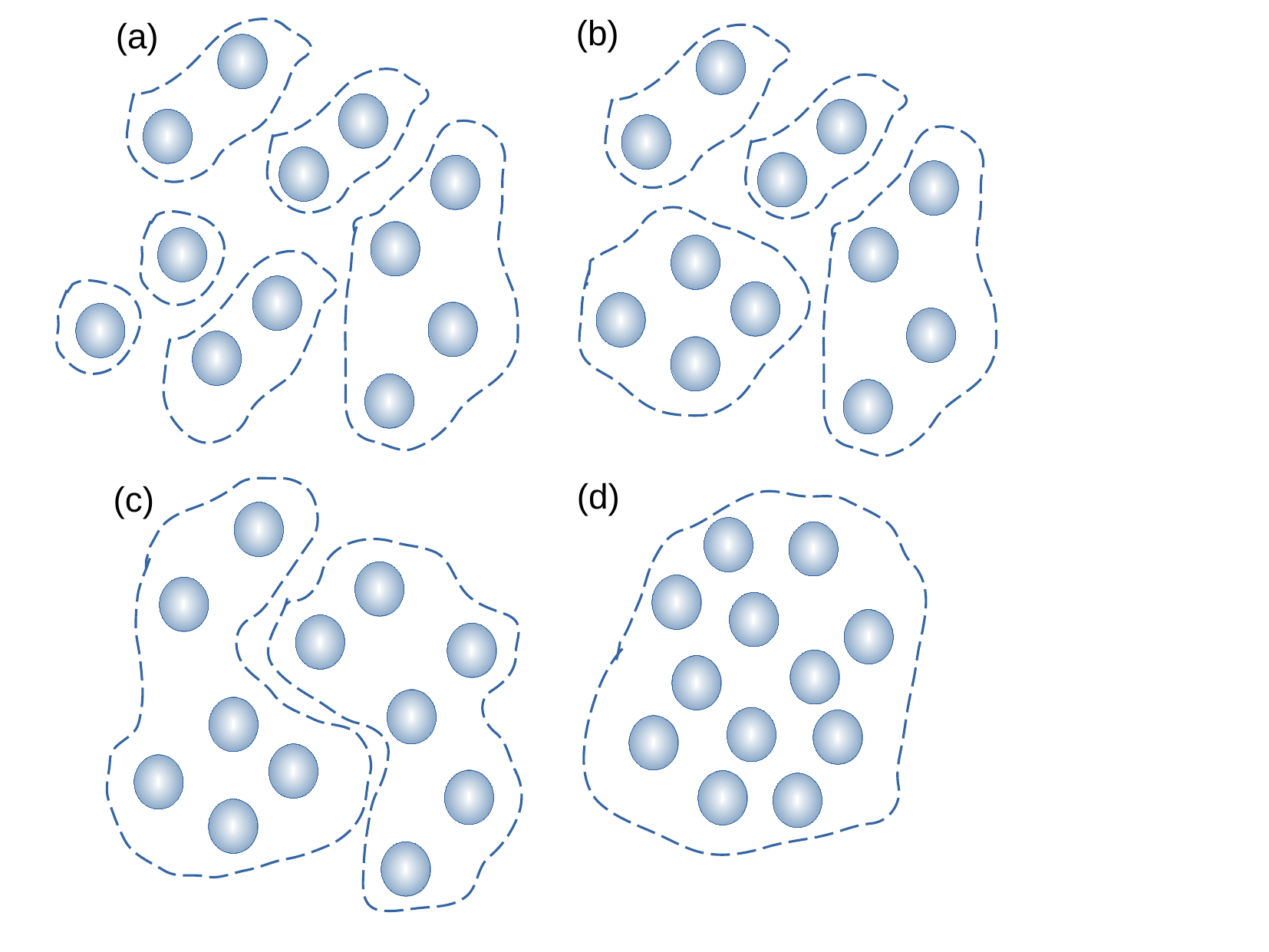}
    \caption{Pictorial representation of $N=12$ qubit quantum states. (a) \emph{6}-separable state  with entanglement depth $d_e=4$, (b) \emph{4}-separable state with entanglement depth $d_e=4$, (c) \emph{2}-separable quantum state with entanglement depth $d_e=6$, and (d) genuinely entangled quantum state with entanglement depth $d_e=12$, not-\emph{2}-separable.}
\label{fig:fig1}
\end{figure}

\subsection{Quantum state structure characterization}
In the following we show how
the correlator $\Q$ carries extensive information on the entanglement of a given state $\hat\varrho$  \cite{spiny.milosz, PhysRevLett.129.250402, PhysRevResearch.6.023050,plodzien2024inherent}.

Let us consider the density operator of $N$ qubits as a product of $k$ entangled states, each with $n_l$ qubits, $\sum_{l=1}^{k} n_l = N$, i.e.,
\begin{equation}
    \hat\varrho = \int d\lambda p(\lambda)\bigotimes_{l=1}^{k}\hat\varrho_{n_l}(\lambda)\equiv \hat\varrho_{n_1 \dots n_k}.
\end{equation}
Such states are called $k$-separable~\cite{10.5555/2011464.2011472,Guhne_2010,PhysRevA.78.032101,Gao2011,Gao_2013,Szalay_2019,HONG2021127347}, and its entanglement depth  is defined as $d_e = \max \{n_1 \dots n_k\}$, see Fig.~\ref{fig:fig1}. 

For all $N$-separable states, $k=N$, $\hat\varrho_{n_1\dots n_k}$, $n_l = 1$, $l\in \{1\dots N\}$,  according to Eqs~\eqref{eq.ent.l} and~\eqref{eq.eqn}, the correlator $\Q$ takes non-positive values
\begin{align}
  \Q\leqslant0,
\end{align}
and violation of this inequality indicates entanglement in the system. On the other hand, the  maximal value is obtained for a \emph{non-2}-separable state $\hat\varrho_{N} = |\psi_{\rm ghz}\rangle\langle\psi_{\rm ghz}|$, being a pure Greenberg-Horne-Zeilinger state, which according to Eq.~\eqref{eq.elq} gives $\mathcal E=1/4$, hence
    $\ket{\psi_{\rm ghz}}=\frac1{\sqrt2}\left(\ket1^{\otimes N}+\ket0^{\otimes N}\right)$, 
 \begin{equation}
    \Q=N-1.
\end{equation}
For states being a product of a single qubit and entangled state of the remaining $N-1$ qubits, $\hat\varrho_{1,N-1}$, 
the correlator is  maximal, when the $N-1$ qubits form a GHZ state, hence again using Eq.~\eqref{eq.elq}, we obtain
\begin{equation}
    \Q\leqslant N-2,
\end{equation} 
and its violation indicates genuine entanglement, i.e. state is \emph{non-2}-separable; the same bound is valid for every \emph{2}-separable state, i.e. $\hat\varrho_{n, N-n}$.

Bounds for $\Q$ can be generalized to states being  a product of $n$ single qubits and $(N-n)$-entangled state,  $\hat\varrho_{n_1\dots n_n, N-n}$, $n_l = 1$, $l\in \{1\dots n\}$:
\begin{align}\label{eq.depth}
    N-1-(n+1)<\Q\leqslant N-1-n,
\end{align}
and the correlator $\Q$ indicates  existence of entanglement depth $d_e = N-n$.

This procedure can be applied to other configurations, for instance for  even number of qubits $N=2M$, when $\hat\varrho$ is a product of $N/2$ pairs of entangled qubits, $\hat\varrho_{n_1\dots n_k}$, $n_l = 2$, $l = \{1\dots N/2\}$, the correlator is bounded from above as
\begin{equation}
    \Q \leqslant \frac{N}{2} = M,
\end{equation}
and its violation certifies existence of entanglement depth $d_e 
 = 3$.

Finally, considering $N = 3 M$ qubits, $\hat\varrho_{n_1\dots n_M}$, $n_l = 3$, $l\in \{1,\dots,M\}$, the bound is
\begin{equation}
    \Q \leqslant \frac{2}{3}N = 2 M,
\end{equation}
and its violation indicates entanglement depth $d_e=4$.

In general, for $k$-separable states the correlator $\Q$ is bounded from above
\begin{align}
    \Q \leqslant N - k,
\end{align}
and violation of this inequality witnesses  a \emph{non-k}-separability, i.e.   $\hat\varrho$  is a product of at most $(k-1)$-entangled states.

\begin{center}
\begin{table}[t!]
  \begin{tabular}{|c |c |c |c |c |}
    \hline
    & $N=3$   & $N=4$ & $N=5$ & $N=6$\\ \hline
    $0<\Q\leqslant1$   & $k < 3$    & $k < 4$ & $k < 5$ & $k < 6$ \\ \hline
    $1<\Q\leqslant2$ & gen. ent. & $k < 3$ & $k < 4$ & $k < 5$\\ \hline
    $2<\Q\leqslant3$ & --- & gen. ent. & $k < 3$ & $k < 4$\\ \hline
    $3<\Q\leqslant4$ & --- & --- & gen. ent. & $k < 3$\\ \hline
    $4<\Q\leqslant5$ & --- & --- & ---  & gen. ent.\\ \hline
  \end{tabular}
  \caption{The values of $\Q$ for $N=3,4,5$, and 6 that signal the \emph{non-k}-separability, i.e., $N$ qubit entangled state can only be represented as product of at most $(k-1)$ entangled states. 
  Minimal \emph{non-k}-separability ($k=2$) signifies genuine entanglement (gen. ent.)}\label{tab.Q} 
\end{table}
\end{center}

As examples, let us consider $N = 5$ and $N=6$. For $N=5$ qubits, the density operator $\hat\varrho$ can take the form   $\hat\varrho_{1,4}$, $\hat\varrho_{2,3}$ or $\hat\varrho_{1,2,2}$. The respective value of the correlator are bounded by: $\Q_{1,4} \leqslant 3$ and its violation certifies genuine entanglement $d_e=5$, $\Q_{2, 3}\le 3 $ and its violation certifies entanglement depth $d_e = 4$;  $\Q_{1,2,2 } \leqslant 2$ and its violation certifies entanglement depth $d_e = 3$. 
For $N=6$, the entangled state $\hat\varrho$ can have the following structure: $\hat\varrho_{1,5}$, $\hat\varrho_{2,4}$, $\hat\varrho_{3,3}$, $\hat\varrho_{2,2,2}$. 
The corresponding bounds are: $\Q_{1,5} \leqslant 4$, $\Q_{2,4} \leqslant 4$, $\Q_{3,3} \leqslant 4$, and $\Q_{2,2,2} \leqslant 3$. 
These observations, adapted to the $N=3,4,5,6$ are summarized in Tab.~\ref{tab.Q}.

\subsection{Extraction of $\Q$}

The correlator $\Q$ can be expressed in terms of a single element of the density matrix
, i.e., 
%${\cal C} \equiv \langle \tilde{0}|^{\otimes N}\hat{R}_k^\dagger(\vec{\theta}_k) \hat\varrho\hat{R}_k(\vec{\theta}_k) |\tilde{1}\rangle^{\otimes N}$
${\cal C} \equiv \langle \tilde{0}|^{\otimes N}\hat\varrho|\tilde{1}\rangle^{\otimes N}$, such that determines the GHZ-type coherence between
all the spins $\ket{\tilde{0}}^{\otimes N}$ up and all down $\ket{\tilde{1}}^{\otimes N}$ in the optimal local bases, namely
\begin{align}\label{eq:Q_form}
  \Q=N+\log_4(\modsq{{\cal C}}).
\end{align}
This observation will be relevant in the forthcoming sections of this work. 
Importantly for its entanglement measure interpretation, $\mathcal{Q}$ is convex in the input states, see Appendix~\ref{app:convexity}.

The freedom of choice  of the orientations $\vec r_k$ is of fundamental importance, as it allows to adapt the correlator to the geometry of the state that is scrutinized~\cite{plodzien2024inherent}. To see this, take two examples
of three-qubit GHZ states, 
\begin{subequations}
  \begin{align}
    &\ket{\psi_{\rm ghz}^{zzz}}=\frac1{\sqrt2}\left(\ket{0_z,0_z,0_z}+\ket{1_z,1_z,1_z}\right)\\
    &\ket{\psi_{\rm ghz}^{xyz}}=\frac1{\sqrt2}\left(\ket{0_x,0_y,0_z}+\ket{1_x,1_y,1_z}\right),
  \end{align}
\end{subequations}
where, for instance, $\ket{0_x}$ denotes the eigenstate of $\hat\sigma_x$ with the spin pointing up. Only the correct choice of the three axes  allows for maximizing the value of the GHZ coherence, i.e. all operators rising the spin in the $z$-direction
in the former, and in the $x$, $y$ and $z$ directions in the latter case, allow to maximize the value of $\Q$. In what follows, we always display results for the correlator, optimized over the choice of the separable basis. 
What is crucial from our perspective of classification of entanglement, the value of correlations is invariant under local unitary operations.
Hence the proposed method for entanglement classification respects SLOCC entanglement classes
\footnote{
The two pure multipartite quantum states $\ket{\psi}$ and $\ket{\phi}$ belong to the same SLOCC  entanglement class if and only if it is possible to convert $\ket{\psi} \mapsto \ket{\phi}$ and $\ket{\phi} \mapsto \ket{\psi}$ with non-zero probabilities using local operations and classical communication, i.e.,  if there exist a set of invertible operators $\hat{O}_k$
(so that the following condition is symmetric) such that 
$ \ket{\phi} = \bigotimes_{k=1}^N\hat{O}_k\ket{\psi}$, where $N$ is the number of constituents.}.

As we stated above, our scheme uses local unitary operations, i.e., local rotations, to transform the state into the optimal basis from the perspective of the GHZ-fidelity. 
A similar setup was already considered from a more general perspective: instead of local unitary operations, the authors of Refs.~\cite{Kaniewski_2016,Li_2019} consider all local operations, because they do not introduce entanglement. 
For such a setup there exist lower bounds~\cite{Kaniewski_2016} on this type of GHZ-fidelity; however, these cannot be used in our case where GHZ-fidelity is necessarily smaller or equal due to a restricted set of operations.

\section{Results}\label{sec:Results}

The number of entanglement classes grows rapidly with the number of parties and the local dimension. 
For pure states of two qubits, there are only two entanglement classes, with a Bell state as an extremal representative of one and separable states forming the other. 
Increasing the number of qubits to three yields two disjoint genuinely multipartite entanglement classes, together forming six different classes.
Interestingly, the growth of the number of classes is not even exponential; already for four qubits, there are infinitely many entanglement classes,  
which can be divided into nine~\cite{Verstraete_2002,Zangi_2017,Vintskevich_2023} or forty-nine groups~\cite{Li_2009,Giordano_2022}. 
The entanglement classification of mixed states, i.e. statistical mixtures of different pure states, is even harder, as each pure state can belong to different entanglement class~\cite{Acin_2001_mixed}.

In the following sections we illustrate the power of our method of entanglement classification for $N=3$ qubits states in three SLOCC classes, $N=4$ qubits states in nine SLOCC classes. Next, we characterize $k$-separability for $N=5,6$ qubits AME states, and Dicke states of an arbitrary size. 

\subsection{Three-qubit entanglement classification}\label{sec:Results_3}
We start by the first non-trivial example which requires classification of entanglement. 
First, we review well-known results concerning classification and then we show the effectiveness of our setup. 

\subsubsection{Pure states}
 
In three-qubit system, there are two types of entanglement: genuine tripartite entanglement and the one in which one qubit is separated from the rest. 
In the genuine entanglement case, there are two inequivalent (under SLOCC) types of states: GHZ and W, defined in Appendix~\ref{sec:tensor_rank}~\footnote{The former class encompasses the majority of states in the Haar-measure sense; a state, drawn at random, will belong to the GHZ class with probability 1.}.
One of the key differences between both genuinely entangled states GHZ and W is their tensor rank, see Appendix~\ref{sec:tensor_rank}.
Therefore, in total there are six classes: two genuine, three two-partite entangled (AB-C, AC-B, BC-A) and one separable class. 

In the following we analyze the correlator $\Q$ for three-qubit SLOCC entanglement classes both for pure and mixed states. To do so, we generate the set of three-qubit quantum states as follows
\begin{itemize}
    \item separable class: $3$ Haar-random unitary matrices of size $2\times 2$ acting on $\ket{000}$, yielding $\hat U_A\otimes \hat U_B \otimes \hat U_C\ket{000}$,
    \item two-party entangled classes: $2$ random unitary matrices: one of size $4\times 4$ and another of size $2\times 2$ to entangle only two subsystems, e.g., $\hat U_{AB}\otimes \hat U_C\ket{000}$,
    \item W class: we apply the parametrization $\ket{\psi} = \lambda_0 \ket{000} + \lambda_1 \ket{100} + \lambda_2 \ket{101} + \lambda_3 \ket{110}$ from Ref.~\cite{Acin_2001,Acin_2001_mixed} with   $\lambda_0, \lambda_1, \lambda_2, \lambda_3$ sampled uniformly,
    \item GHZ class: a random state belongs to this class with probability 1~\cite{Acin_2001}; therefore, we act with a random unitary matrix of size $8\times 8$ on any pure state: $\hat U_{ABC}\ket{000}$.
\end{itemize}

\begin{figure}[t!]
\includegraphics[width=.8\linewidth]{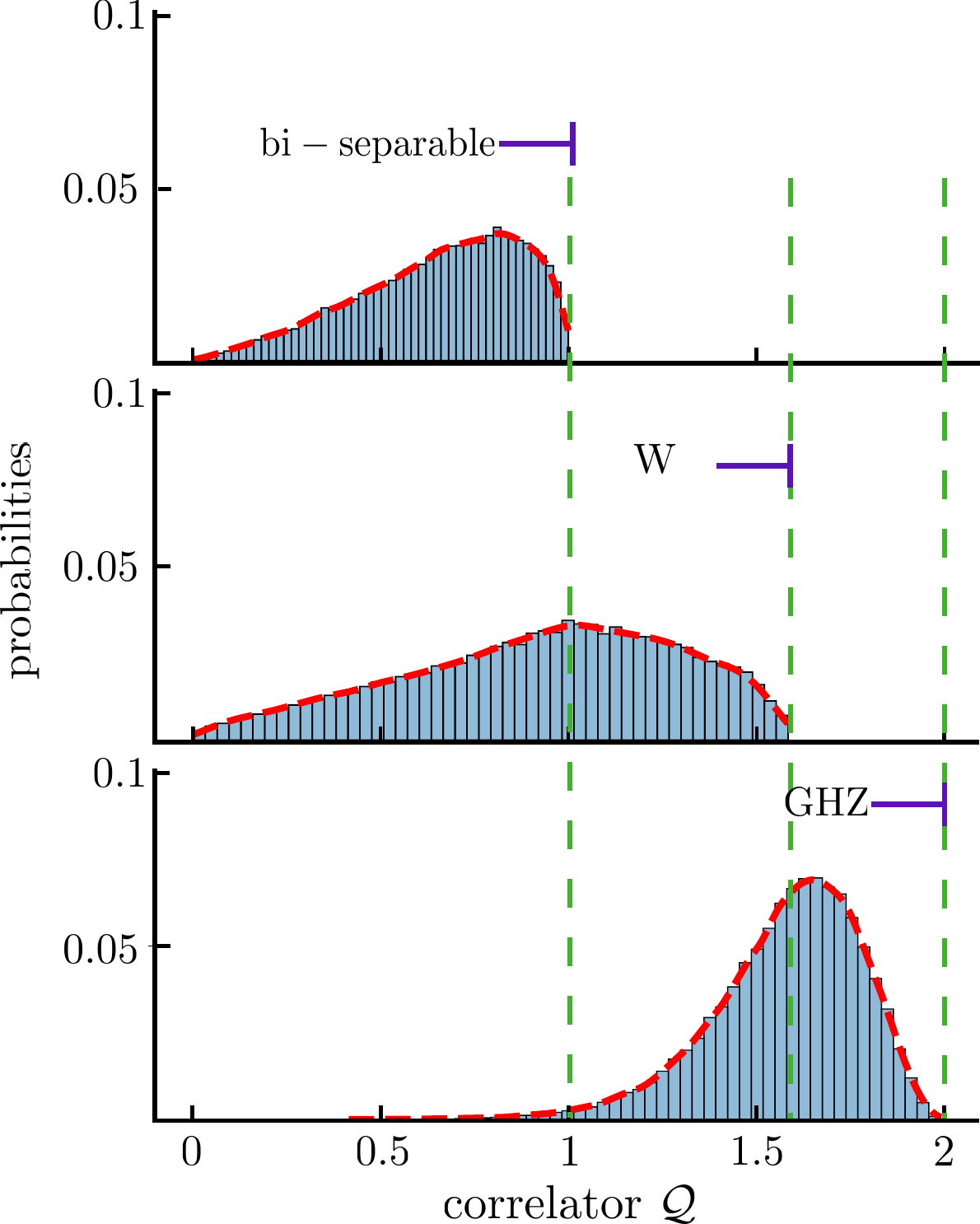}
    \caption{Probability distribution of $\Q$ for three-qubit pure states from SLOCC classes: bi-separable-, W-, and GHZ-type (from top to bottom).
    The green dashed lines correspond to the maximal values of $\Q$ for each class. Red dashed lines presents numerically obtained probability distribution function. {Dimensionless units.}}
    \label{fig:fig2}
\end{figure}

 We generate $100$ states of a maximal representatives of each class, and $20000$ states for non-maximal representatives of each class. Next, we calculate the optimized Bell correlator $\Q$ for each state.  

In Fig.~\ref{fig:fig2} we show the distribution of outcomes, separately for each class.
First, we focus on the maximal representatives of each class. In this case, correlator $\Q$ takes always same value, i.e. $\Q=1$ for bi-separable states, $\Q = 1.6$ for W states, and $\Q = 2$ for GHZ states -- denoted as vertical green lines in Fig.~\ref{fig:fig2}. This clearly indicates that correlator $\Q$ does not depend on local rotations, and distinguish between SLOCC classes properly.
The bound $\Q\leqslant 1.6$ serves as a GHZ-class witness. When $\Q>1.6$ for either pure or mixed states, state $\hat\varrho$ does not belong to the W-class.
Values of the correlator $\Q$ indicates the genuine entanglement of states W and GHZ, in accordance with Tab.~\ref{tab.Q}.

Next, let us consider the value of $\Q$ for non-maximal representatives of each class. Frequencies of $\Q$ for each class are presented as histogram in Fig.~\ref{fig:fig2}. Interestingly, for class W the most probable is $\Q=1$. For the GHZ class the most probable is $\Q = 1.65$. 
From the purely numerical results, we observe  that most probable value of $\Q$ for W class is the the maximal value of $\Q$ for bi-separable class, while most probable value of $\Q$ for GHZ class is close to the maximal value of $\Q$ for the W class.
 
Finally, the following question arises: how precisely does $\Q$ distinguish between the classes. To address this point we 
use the Wootters' distance, which for probability distributions $p$ and $q$ of a random variable $n$ is defined as
\begin{align}
  d(p,q)=1-\sum_n\sqrt{p_nq_n},
\end{align}
where the measure $\sum_n\sqrt{p_nq_n}$ is known as the Bhattacharyya coefficient.
Two probabilities are distant if $d(p,q)\simeq1$ and naturally $d(p,p)=0$.
We calculate the Wootters' distance, pairwise, for the probabilities shown in Fig~\ref{fig:fig2} (red lines). The outcome, shown in Table~\ref{tab:Wootters_dist_3q}, informs about the success chance of discrimination 
between the classes. The GHZ and the bi-separable states are the most distant, confirming the result of Fig.~\ref{fig:fig2}. As expected, the W-class that lies in between the two has a significant overlap with the both.

To assign a state to one of  {  three} given classes one can process it through the random SLOCC many times,
calculating the correlator for each case. This procedure will give  one of the   {  three} shapes from Fig.~\ref{fig:fig2}. The degree of distinguishability can be controlled with the number of samples.
This protocol can be thought of as a black-box device, with random SLOCC class states on the input that, as an output, produces the correct classification provided sufficiently many samples.

\begingroup
\renewcommand*{\arraystretch}{1.5}
\begin{table}[t!]
  \begin{tabular}{|c ||c |c| c|}
    \hline
   Ent. class & B-S  & W  & GHZ  \\ \hhline{|=||=|=|=|}
    B-S   & 0    & 0.27  & 0.93 \\ \hline
    W  & 0.27    & 0  & 0.57\\ \hline
    GHZ   & 0.93   & 0.57  & 0\\ \hline
\end{tabular}
\caption{The table of pairwise Wootters' distance between the probability distributions of the optimized correlators $\Q$ for three-qubit entanglement classes: biseparable (B-S), W, and GHZ.}\label{tab:Wootters_dist_3q}
\end{table}
\endgroup

\subsubsection{Mixed states}
 
\begin{figure}[t]
\includegraphics[width=1\linewidth]{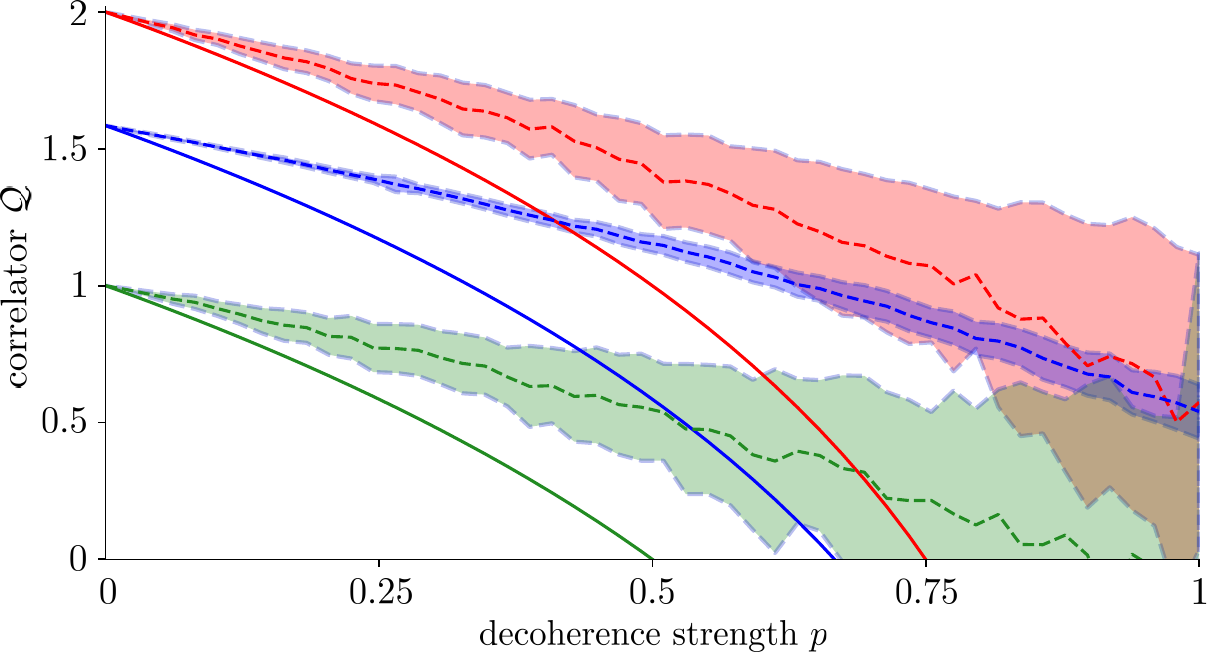}
    \caption{The value of the correlator $\Q$ for extremal representatives from each of three-qubit entanglement class, upon which we act with dephasing (dashed lines) and depolarizing (solid lines) noise. 
    The red lines correspond to the noisy GHZ states, the blue one to W states, and the green line depicts the decay of correlations for two-qubit Bell states.  {Dimensionless units.}}
    \label{fig:fig4}
\end{figure}

In the following, we study how $\Q$ is vulnerable to noise, i.e., a given state $\ket\psi$ with a given value of $\Q$ belonging to some entanglement class, is passed through a decoherence channel.
We take into account only states which give the maximal value of $\Q$ for each class (see the vertical dashed green lines in Fig.~\ref{fig:fig2}).
We consider two types of decoherence, where the first represents a global depolarizing channel 
\begin{align}\label{eq:depolarizing_channel}
  \Lambda_p[\hat{\varrho}] = (1-p)\hat{\varrho}+\frac{p}{8}\hat{\mathds1}.
\end{align}
As it does not depend on the orientation of the optimal correlator, the value of $\Q$ is analytical and reads
\begin{equation}
  \Q(p) =\Q+2\log_4(1-p).
\end{equation}
This result is drawn in  Fig.~\ref{fig:fig4} with solid lines. 

The other source of noise is the dephasing acting on randomly chosen qubit $i \in \{1,2,3\}$ 
\begin{equation}\label{eq:channel_dephasing}
  \Delta_p[\hat{\varrho}] = (1-\frac{p}{2})\hat{\varrho} + \frac{p}{2}\hat{\sigma}^{(i)}_z\hat{\varrho}\hat{\sigma}^{(i)}_z.
\end{equation}
 For every $p$ we generate  $100$ random states of maximal representatives of each class, and  we obtain three values of the optimized $\Q$, depending on which qubit the dephasing acts. These samples
give a range of $\Q$'s, which together with the average value is shown as a shaded area for each class separately, see Fig.~\ref{fig:fig4}.
We observe that for reasonably small noise levels ($p < 0.3$ for the dephasing and $p < 0.25$ for the depolarizing noise), the maximal representatives of the SLOCC entanglement classes for three qubits are still mutually distinguishable using the Bell correlators metric.

\subsubsection{Application: dynamical generation of entanglement}

\begin{figure}[t!]
\includegraphics[width=1\linewidth]{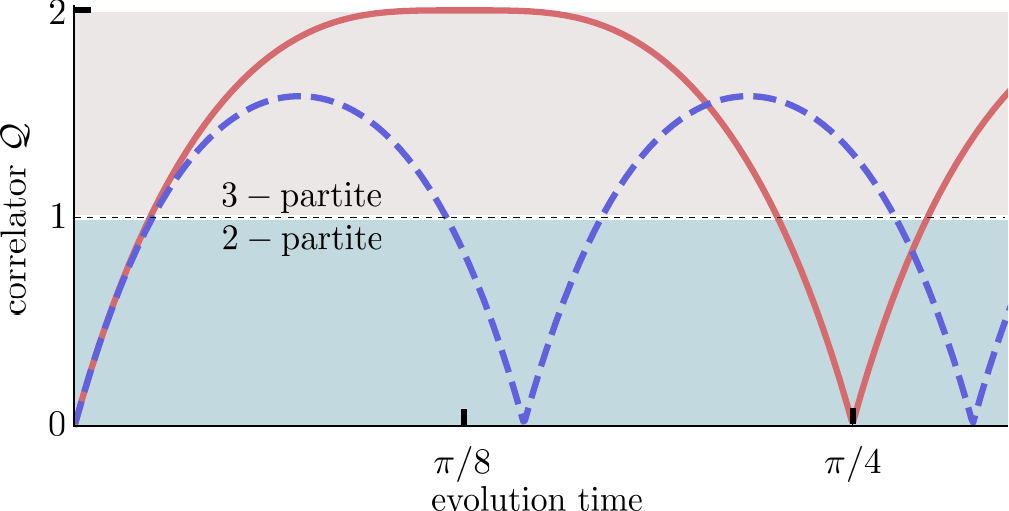}
    \caption{Dynamical generation of the many-body entanglement in spin squeezing protocols in three-qubit system, i.e., $\Q$ versus the time $t$. The red solid line corresponds to the one-axis twisting protocol (OAT), while the blue dashed line corresponds to the two-axis counter twisting protocol (TACT). Both protocols generate genuine entanglement at the same critical time $t_c$, when $\Q(t_c)$ exceeds unity. The OAT protocol generates the most entangled state, namely the GHZ state, for which the correlator takes the maximal value $\Q = 2$.  {Dimensionless units.}}
    \label{fig:fig3}
\end{figure}

Next, we apply this correlator to classify the three-qubit entangled states generated in the following dynamical protocol.
Qubits are initialized in the product state of eigenstate of the $\hat\sigma_x^{(k)}$ operators, namely
\begin{align}
  \ket{\psi(0)}=\left(\frac{\ket0+\ket1}{\sqrt2}\right)^{\otimes3}
\end{align} 
and undergo the evolution
\begin{align}
  \ket{\psi(t)} = e^{-it\hat H }\ket{\psi(0)}
\end{align}
generated by either of the two Hamiltonians
\begin{subequations}
  \begin{align}
    \hat{H}_{\rm oat}&=\sum_{i>j=1}^3\hat{\sigma}^{(i)}_z\hat{\sigma}^{(j)}_z\\
    \hat{H}_{\rm tact}&=\sum_{i>j=1}^3\left(\hat{\sigma}^{(i)}_z\hat{\sigma}^{(j)}_z-\hat{\sigma}^{(i)}_x\hat{\sigma}^{(j)}_x\right).
  \end{align}
\end{subequations}
Here, the two subscripts stand for one-axis- or two-axis-counter-twisting, respectively. 
 Note that according to Eq.~\eqref{eq.depth} and the Tab.~\ref{tab.Q} both these Hamiltonians generate entanglement at all times, and genuinely 3-partite entanglement at most instants, see  Fig.~\ref{fig:fig3}.
  Moreover, the correlator $\Q$ provides information that only the OAT protocol generates the GHZ-class state with $\Q=2$ at maximum.  

\subsection{Four-qubit entanglement classification}\label{sec:Results_4}

The  efforts to classify the multipartite entanglement, relevant in the context of experimental advances~\cite{Sackett2000},
have been  recently expanded to  the four-qubit case~\cite{Ghahi_2016,Vintskevich_2023,Wieczorek_2009,Kiesel_2007}, where nine groups can be distinguished~\cite{Verstraete_2002}.  The main group is formed by those states that are characterized by the generalized GHZ-type entanglement. The other groups are formed by three- and two-qubit entangled states.

For each of the nine classes, we maximize the correlator $\Q$ by picking  the maximally entangled representatives of each group (Appendix~\ref{app:4_qubit_classes} and Ref.~\cite{Vintskevich_2023}).
As shown in Table~\ref{tab:4-qubit_classes}, the highest value of the correlator is reached by the GHZ state ($|G\rangle$ group), and $|E_1\rangle$ and $|E_2\rangle$ groups, while the smallest $\Q$ is for group $|E_6\rangle$.

An important question is about $k$-separability of a resulting mixed quantum state after tracing out one of the qubits. The correlator $\Q$ provides an information about $k$-separability of the final three-qubit state.
When $\Q>0$ the remaining quantum state is entangled. 
Naturally, the  entanglement in states $|G\rangle$ is washed out by tracing out any of the qubits. This however is not the case with classes $\ket{E_3}$, $\ket{E_4}$, $\ket{E_6}$, $\ket{E_7}$, and $\ket{E_8}$, see Table~\ref{tab:4-qubit_classes}.
Moreover, when   $\Q>1$, the mixed states remains genuinely entangled. In particular state $|E_7\rangle$ remains genuinely entangled after tracing out the first qubit, $\Q = 1.32$.

\begin{table}[t!]
  \begin{tabular}{|c |l ||l |l |l |l |}
    \hline
    \diagbox[height=3.3em,width=6.3em]{State}{Discarded}  & None  & $i = 1$  & $i = 2$   & $i=3$  & $i=4$  \\ \hline
    
    $|G\rangle$   & 3    & $-$  & $-$   & $-$  & $-$  \\ \hline
$|E_1\rangle$ & 2.98 & $-$  & $-$   & $-$  & $-$  \\ \hline
$|E_2\rangle$ & 2.98 & $-$  & $-$   & $-$  & $-$  \\ \hline
$|E_3\rangle$ & 2.41 & $-$  & $-$   & $0.43$  & $0.43$  \\ \hline
$|E_4\rangle$ & 2.5 & $0.55$  & $-$   & $0.68$  & $-$  \\ \hline
$|E_5\rangle$ & 2.98 & $-$  & $-$   & $-$  & $-$  \\ \hline
$|E_6\rangle$ & 1.54 & $1$  & $-$   & $-$  & $1$  \\ \hline
$|E_7\rangle$ & 1.76 & $1.32$  & $1$   & $1$  & $1$  \\ \hline
$|E_8\rangle$ & 2 & $2$  & $-$   & $-$  & $-$  \\ \hline
\end{tabular}
  \caption{Value of the correlator $\Q$ for maximally entangled four-qubit states in nine entanglement classes (see Appendix~\ref{app:4_qubit_classes}). The first column corresponds to the value of the correlator of the full state, while the latter four show $\Q$ after discarding (tracing out) one qubit.
    Null values corresponds to negative lack of entanglement. 
    Observe that the state exhibiting the highest entanglement with respect to four-qubit correlator ($\ket{G}$) does not possess any three-qubit entanglement. 
    Conversely, the states with lower four-qubit entanglement preserve high values of the three-qubit correlator.
    This is in agreement with similar observations concerning distribution of entanglement among different divisions~\cite{Verstraete_2002}. }\label{tab:4-qubit_classes}
\end{table}

\subsection{non-$k$-separability of AME states for $N=5$ and $6$ qubits}\label{sec:Results_AME}

A pure state on $N$ qu$d$its is called an \textit{absolutely maximally entangled}  AME($N$,$d$) state if all its reduced density matrices after discarding at least half of the subsystems are maximally mixed~\cite{Helwig_2012}. 
Interestingly, the question of the existence of an AME state of a given number of parties and local dimensions is still open. 
There are a few methods for (dis)proving their existence but none is universal~\cite{Table_AME,Scott_2004}. 
Recently there were several developments in the search for AME states, including the question of their local orbit, connected with local unitary equivalence~\cite{Huber_2017,Rather_2022,Rather_2023}. The usefulness of AME states extends also to dense coding, quantum secret sharing, and quantum error correction codes~\cite{Scott_2004,Goyeneche_2015}.

Here, we consider only the qubit case, $d=2$. 
In the case of $N=2,3$ qubits, the AME  {state} is the corresponding GHZ state.
Surprisingly, there are no four-qubit AME states~\cite{Higuchi_2000}, and
there are no AME states for  $N>6$~\cite{Scott_2004,Huber_2017}; 
but they do exist  
for $N=5, 6$ \cite{Facchi_2008,Goyeneche_2015}, taking the form of 
\begin{align}
  \ket{\text{AME}(N,2)} = \frac{1}{2^{N/2}} \sum_{i=0}^{2^N-1}\alpha^{(N)}_i \ket{i},
\end{align}
where the values of binary ($\pm1$) coefficients $\alpha^{(5)}_i$ and $\alpha^{(6)}_i$ are given in Appendix~\ref{app:AME}.
For ${\rm AME}(5,2)$ the correlator takes $\Q = 1.7$ and for ${\rm AME}(6,2)$ it takes value $\Q = 2.4$. These values are about half of maximal values for $N = 5$ and $N = 6$ qubits ($\Q = 4$ and $\Q = 5$ respectively,  see Table~\ref{tab.Q}.).

The type of entanglement of AME states cannot be explained via the provided measure; the correlator for those two AME states can be explained with only three-partite and four-partite correlations, respectively. 
In this sense, AME's are the strongest entangled states that exist.
The correlator $\Q$ can be thus thought of as an alternative to the usual description of AME-type entanglement, understood as the sum of entanglement entropies across every balanced bipartition.

\subsection{non-$k$-separability of Dicke states, any $N$}\label{sec:Results_Dicke}

As the last illustration of $k$-separability characterization with the correlator $\Q$ we use the Dicke states. A set of $N$-qubit Dicke states $|J, m\rangle$ is defined as a eigenbasis of collective multi-qubit spin operators $\hat{J}^2 = \hat{J}^2_x + \hat{J}^2_y +\hat{J}^2_z$ and $\hat{J}_z$, where $\hat{J}_\tau = \frac{1}{2}\sum_{i=1}^N\hat{\sigma}^{(i)}_\tau$, $\tau = x,y,z$;
$[\hat{J}^2, \hat{J}_z] = 0$, 
\begin{equation}
    \begin{split}
        \hat{J}^2|J,m\rangle& = J(J+1)|J,m\rangle \\
        \hat{J}_z|J,m\rangle & = m|J,m\rangle,
    \end{split}    
\end{equation}
with $J = N/2$, $m\in\{-J\dots J \}$.
The special case are balanced Dicke states, $m = 0$,
\begin{equation}
    |D_n\rangle = \binom{N}{\frac{N}{2}}^{-\frac{1}{2}}\sum_k {\cal P}_k(|0\rangle^{\otimes \frac{N}{2}}\otimes|1\rangle^{\otimes \frac{N}{2}}),
\end{equation}
summed over all distinct permutations ${\cal P}_k$.
Such symmetric Dicke states posses strong multipartite entanglement \cite{PhysRevLett.112.155304, Campbell_2009,  PhysRevLett.107.180502,  PhysRevA.94.042333, Demianowicz_2021, Aloy_2021, Marconi2021entangledsymmetric, Tura2018separabilityof, PhysRevA.95.042128,PhysRevA.104.022426}, and are useful for quantum metrology \cite{Toth:07, hyllus2012fisher, toth2012multipartite}.

For Dicke states the correlator $\Q$ can be obtain analytically \cite{plodzien2024inherent} both for balanced $m=0$ case, and for $m\ne0$:
\begin{equation}
    \Q = 2\log_4\binom{N}{m+\frac{N}{2}}.
\end{equation}
For the large number of qubits,  $N\to\infty$, the above formula can be approximated as 
\begin{equation}\label{eq:Q_Dicke}
    \Q = N - \log_4N-4\frac{m^2}{N}\log_4 e + \log_4\frac{2}{\pi}.
\end{equation}
The maximal value of the correlator $\Q$ is for balanced Dicke states with $m=0$.
 Moreover, the   expression in Eq.~\eqref{eq:Q_Dicke} and inequality Eq.~\eqref{eq.depth} allow classification of  Dicke states $|J,m\rangle$ with respect to their \emph{non-k}-separability, one of the results of this work.

\section{Measurements}\label{sec:Measurement}

As discussed in the previous section, many-qubit entanglement can be inferred from the single element of the density matrix $\hat{\varrho}$ related to GHZ fidelity.
 The remaining question concerns the identification of an experimental protocol capable of accessing this quantity. 

The measurement of the off-diagonal GHZ element of the density matrix in $N=4$ qubit system has been performed experimentally with trapped ions \cite{Sackett2000}. Another approach is to use the quantum state tomography, e.g., with classical shadows, and rotate the state to find the optimal basis in 
which the GHZ coherence element is maximal~\cite{PhysRevResearch.6.023050}. 
An alternative experimental protocol, not relying on the full tomography, is based on the multiple quantum coherences measurements. This technique provides extensive information about the structure of a many-body state, and allows intertwining of the quantum correlators with other physical quantities like the out-of-time-order correlations~\cite{garttner2017measuring,PhysRevLett.120.040402}.

\section{Conclusions}\label{Sec:Conclusions}

In this work we have provided a many-body quantum correlator that captures the \emph{non}-$k$-separability of a given pure or mixed quantum state. By construction, the proposed correlator respects the SLOCC entanglement classification; we used it to classify three entanglement classes in a three-qubit system, and to characterise nine entanglement classes in a four-qubit system. We then characterised \emph{non}-$k$-separability for qubit AME states, and Dicke states of an arbitrary number of qubits. The proposed method is not limited by the system size, and is experimentally accessible.

Like any other method, the one presented in this paper is not able to perfectly classify all entangled states. 
The strength of the method lies rather in its simplicity. 
This comes at a price as for each state one has to choose a suitable basis by local unitary rotations, such that maximises the generalised GHZ fidelity. 

Our setup is not specifically designed for pure states, and can be adapted to mixed states as well.  Therefore, the technique is able to statistically distinguish between states even in the more realistic, mixed case scenarios with experimental imperfections, including particle losses.
In many experimental setups it is not necessary to study the most general form of entanglement possible but rather to confirm that the losses do not hinder the state preparation. 
For such setups, our framework fits perfectly. 

\section*{Acknowledgement}
We thank Guillem M\"uller-Rigat, Anna Sanpera, Geza Toth, and Karol Życzkowski for reading the manuscript and their useful comments.
JCh was supported by the National Science Centre, Poland, within the QuantERA II Programme that has received fund
ing from the European Union’s Horizon 2020                      
research and innovation programme under Grant Agreement No 101017733, Project No. 2021/03/Y/ST2/00195.
ICFO group acknowledges support from:
European Research Council AdG NOQIA; MCIN/AEI (PGC2018-0910.13039/501100011033, CEX2019-000910-S/10.13039/501100011033, Plan National FIDEUA PID2019-106901GB-I00, Plan National STAMEENA PID2022-139099NB, I00, project funded by MCIN/AEI/10.13039/501100011033 and by the “European Union NextGenerationEU/PRTR" (PRTR-C17.I1), FPI); 
QUANTERA MAQS PCI2019-111828-2; QUANTERA DYNAMITE PCI2022-132919, QuantERA II Programme co-funded by European Union’s Horizon 2020 program under Grant Agreement No 101017733; Ministry for Digital Transformation and of Civil Service of the Spanish Government through the QUANTUM ENIA project call - Quantum Spain project, and by the European Union through the Recovery, Transformation and Resilience Plan - NextGenerationEU within the framework of the Digital Spain 2026 Agenda; Fundació Cellex; Fundació Mir-Puig; Generalitat de Catalunya (European Social Fund FEDER and CERCA program, AGAUR Grant No. 2021 SGR 01452, QuantumCAT \ U16-011424, co-funded by ERDF Operational Program of Catalonia 2014-2020); Barcelona Supercomputing Center MareNostrum (FI-2023-3-0024);  Funded by the European Union. 

Views and opinions expressed are however those of the author(s) only and do not necessarily reflect those of the European Union, European Commission, European Climate, Infrastructure and Environment Executive Agency (CINEA), or any other granting authority.  Neither the European Union nor any granting authority can be held responsible for them (HORIZON-CL4-2022-QUANTUM-02-SGA  PASQuanS2.1, 101113690, EU Horizon 2020 FET-OPEN OPTOlogic, Grant No 899794),  EU Horizon Europe Program (This project has received funding from the European Union’s Horizon Europe research and innovation program under grant agreement No 101080086 NeQSTGrant Agreement 101080086 — NeQST);  ICFO Internal “QuantumGaudi” project;  European Union’s Horizon 2020 program under the Marie Sklodowska-Curie grant agreement No 847648; “La Caixa” Junior Leaders fellowships, La Caixa” Foundation (ID 100010434): CF/BQ/PR23/11980043.

\appendix

\section{Convexity of correlator $\mathcal{Q}$}\label{app:convexity}
 
Entanglement cannot be increased while mixing the quantum states~\cite{PhysRevLett.78.2275,Horodecki2001,Plenio_2007}, therefore, any entanglement measure should be convex. As such we require that  for any two states $\hat\varrho_1$, $\hat\varrho_2$, the correlator $\Q$ fulfills
\begin{equation}\label{eq:convexity}
    \Q(\lambda \hat\varrho_1 + (1-\lambda)\hat\varrho_2) \leqslant \lambda\Q(\hat\varrho_1) + (1-\lambda)\Q(\hat\varrho_2).
\end{equation}

The correlator $\mathcal{Q}$, Eq.\eqref{eq:Q_form}, for a linear combination of two states reads:
\begin{equation}
    \begin{split}
        \mathcal{Q}\big(\lambda &\hat\varrho_1 + (1-\lambda) \hat\varrho_2\big) = \\
        &= N + \log_4 |\langle \tilde{0}|^{\otimes N}(\lambda \hat\varrho_1 + (1-\lambda) \hat\varrho_2)|\tilde{1}\rangle^{\otimes N}|^2.       
    \end{split}
\end{equation}

Let us now focus on the logarithm. 
It inherently involves maximization over a choice of basis, which can be upper bounded by two independent optimizations
\begin{equation}
    \begin{split}
        2 \log_4\bigg(\max_{\{\ket{\tilde{0}},\ket{\tilde{1}}\}} &|\langle \tilde{0}|^{\otimes N}(\lambda \hat\varrho_1 + (1- \lambda) \hat\varrho_2)|\tilde{1}\rangle^{\otimes N}|\bigg) \leqslant \\
        \leqslant 2  \log_4 \bigg(&\lambda\max_{\{\ket{\tilde{0}},\ket{\tilde{1}}\}} |\langle \tilde{0}|^{\otimes N}\hat\varrho_1 |\tilde{1}\rangle^{\otimes N}| \,\,\, + \\
        &+ (1-\lambda)\max_{\{\ket{\tilde{0}},\ket{\tilde{1}}\}} |\langle \tilde{0}|^{\otimes N}\hat\varrho_2 |\tilde{1}\rangle^{\otimes N}|\bigg).
    \end{split}
\end{equation}
Using the concavity of the logarithm, we can bound the latter expression by
\begin{equation}
    \begin{split}
        \lambda \log_4\bigg(\max_{\{\ket{\tilde{0}},\ket{\tilde{1}}\}} &|\langle \tilde{0}|^{\otimes N}\hat \varrho_1 |\tilde{1}\rangle^{\otimes N}|^2\bigg) + \\ 
        + &(1-\lambda) \log_4\bigg(\max_{\{\ket{\tilde{0}},\ket{\tilde{1}}\}} |\langle \tilde{0}|^{\otimes N}\hat\varrho_2 |\tilde{1}\rangle^{\otimes N}|^2\bigg),
    \end{split}
\end{equation}
from which we directly obtain the expressions for $\mathcal{Q}(\hat\varrho_1)$ and $\mathcal{Q}(\hat\varrho_2)$, yielding Eq.~(\ref{eq:convexity}), what proves the convexity of the correlator $\Q$.

\section{Tensor rank and multipartite entanglement classification}\label{sec:tensor_rank}
For two-particle pure states
$\ket\psi = \sum_{i_1i_2} \alpha_{i_1i_2} \ket{i_1i_2}$,
classification and quantification of bipartite pure entanglement stems from the singular value decomposition (SVD). This procedure allows to tansform
the matrix of the coefficients $[\hat\alpha]_{i_1i_2} = \alpha_{i_1i_2}$ to give the Schmidt decomposition of the state $\ket\psi = \sum_i \sqrt{\lambda_i} \ket{\chi_i,\xi_i}.$
From the vector of the singular values $\Lambda = [\lambda_i]_i$ it is possible to infer all the information about entanglement, such as the entanglement entropy$-\sum_i \lambda_i \ln{\lambda_i}$, or the Schmidt rank (number of non-zero $\lambda_i$s)~\cite{Bengtsson_2006}. 
For example, one can convert one two-partite state $\ket\psi$ into another $\ket\phi$ via SLOCC if and only if the Schmidt rank of the latter is not larger.

When $N\geqslant3$, the situation is more complicated because the coefficients of the state $\ket{\psi} = \sum_{i_1\ldots i_N}\alpha_{i_1\ldots i_N}\ket{i_1\ldots i_N}$,
form a tensor, for which, in contrast to matrices, SVD is not always possible
and various generalizations of higher-order SVDs were proposed (for a list of their applications in quantum information see Ref.~\cite{Bruzda_2023}).
One way is to use the \textit{tensor rank}, which is the minimal number of separable states into which one can decompose a given state. 
We use this notion to exemplify the difference between two \emph{most entangled} states of three qubits, namely between 
$\ket{\text{GHZ}} = \frac{1}{\sqrt{2}}(\ket{000}+\ket{111})$, which has a rank 2 and $\ket{\text{W}} = \frac{1}{\sqrt{3}}(\ket{001} + \ket{010} + \ket{100})$, which has a rank 3. 

Surprisingly, there are states of tensor rank 2 that approximate the W state to an arbitrary precision~\cite{Acin_2001, Vrana_2015}, but the converse is not true -- the GHZ state does not have states of tensor rank 3 in its neighborhood inside the Hilbert space.
It is again in direct contrast with the bipartite case, as matrices of lower rank always have matrices of higher rank in the neighborhood, while the converse does not hold. 
Therefore, we say that the \textit{border rank} of the W state is $2$. 
What is more, the generic rank of a matrix is the maximal one, but for tensors of three indices 2-level each (so three-qubit pure states) this is not the case; the maximal tensor rank is 3 and the generic is 2.
As such, the generic pure state of three qubits belongs to the GHZ SLOCC-class, not to the W class. 

\section{Entanglement classes for four qubits}\label{app:4_qubit_classes}
Here we list nine representatives of entanglement classes for four qubits \cite{Verstraete_2002, Zangi_2017,Vintskevich_2023}:
\begin{widetext}
\begin{equation}
    \begin{split}
        \ket{G} &= \frac{a+b}{2} \big( \ket{0000} + \ket{1111} \big) + \frac{a-d}{2}\big( \ket{0011} + \ket{1100} \big) + \frac{b+c}{2}\big( \ket{0101} + \ket{1010} \big) + \frac{b-c}{2}\big( \ket{0110} + \ket{1001} \big), \\ 
        \ket{E_1} &= \frac{a+b}{2} \big( \ket{0000} + \ket{1111} \big) + \frac{a-b}{2}\big( \ket{0011} + \ket{1100} \big) + c\big( \ket{0101} + \ket{1010} \big) + \ket{0110}, \\ 
        \ket{E_2} &= a \big( \ket{0000} + \ket{1111} \big) + b \big( \ket{0101} + \ket{1010} \big) + \ket{0110} + \ket{0011}, \\
        \ket{E_3} &= a \big( \ket{0000} + \ket{1111} \big) + \frac{a+b}{2} \big( \ket{0101} + \ket{1010} \big) + \frac{a-b}{2} \big( \ket{0110} + \ket{1001} \big) + \frac{i}{\sqrt{2}} \big( \ket{0001} + \ket{0010} + \ket{0111} + \ket{1011} \big), \\ 
        \ket{E_4} &= a \big( \ket{0000} + \ket{0101}  + \ket{1010} + \ket{1111} \big) + i\ket{0001} + \ket{0110} - i\ket{1011}, \\
        \ket{E_5} & = a\big(\ket{0000} + \ket{1111}\big) + \ket{0011} + \ket{0101} + \ket{0110},\\
        \ket{E_6} & = \ket{0000} + \ket{0101} + \ket{1000} + \ket{1110}, \\
        \ket{E_7} & = \ket{0000} + \ket{1011} + \ket{1101} + \ket{1110}, \\
        \ket{E_8} & = \ket{0000} + \ket{0111}.
    \end{split}
\end{equation}
\end{widetext}

 \section{Absolutely Maximally Entangled states AME(5,2) and AME(6,2)}\label{app:AME}
The AME coefficients \cite{Facchi_2008,Goyeneche_2015} for $N=5$ are
\begin{equation}
\begin{split}
    \alpha^{(5)} & = \{1, 1, 1, 1, 1, -1, -1, 1, 1, -1, -1,  1, \\& 
                       1, 1, 1,  1, 1, -1, -1,  1, -1, 1, -1, -1, \\& 1, -1, 1, -1, -1, 1, 1\}
\end{split}
\end{equation}

while for $N=6$
\begin{equation}
  \begin{split}
    \alpha^{(6)} & = \{-1, -1, -1, 1, -1, 1, 1, 1, -1, -1, -1, 1,     \\&
                    1, -1, -1, -1, -1, -1, 1, -1, -1, 1, -1, -1,     \\&
                    1, 1, -1, 1, -1, 1, -1, -1, -1, 1, -1, -1, -1,   \\& 
                    -1, 1, -1, 1, -1,1, 1, -1, -1, 1, -1, 1, -1, -1, \\& 
                    -1, 1, 1, 1, -1, 1, -1, -1, -1, -1, -1, -1, 1\}. 
\end{split}
\end{equation}

\bibliography{biblio}

\end{document}